# BUG SEVERITY PREDICTION IN SOFTWARE PROJECTS USING SUPERVISED MACHINE LEARNING MODELS

**BY**

Nafisha Tamanna Nice
ID: 241-25-015

This Report Presented in Partial Fulfillment of the Requirements for the Degree of Masters of Science in Computer Science and Engineering

**Supervised By**
**Professor Dr. Sheak Rashed Haider Noori**
Professor and Head
Department of CSE
Daffodil International University

**Co-Supervised By**
**Dr. Abdus Sattar**
Associate Professor & Director
Department of CSE
Daffodil International University

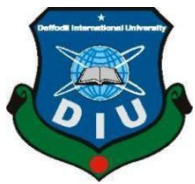

**DAFFODIL INTERNATIONAL UNIVERSITY**

**DHAKA, BANGLADESH**

**DECEMBER 2025**

# APPROVAL

This Thesis titled "**Bug Severity Prediction in Software Projects Using Supervised Machine Learning Models**", submitted by **Nafisha Tamanna Nice,** ID No: **241-25-015** to the Department of Computer Science and Engineering, Daffodil International University, has been accepted as satisfactory for the partial fulfillment of the requirements for the degree of M.Sc. in Computer Science and Engineering (MSc.) and approved as to its style and contents. The presentation has been held on 10/01/2026.

## **BOARD OF EXAMINERS**

\_\_\_\_\_\_\_\_\_\_\_\_\_\_\_\_\_\_\_\_
**(Name)**                                                                                                                    **Chairman**
**Designation**
Department of Computer Science and Engineering
Faculty of Science & Information Technology
Daffodil International University

\_\_\_\_\_\_\_\_\_\_\_\_\_\_\_\_\_\_\_\_
**(Name)**                                                                                                           **Internal Examiner**
**Designation**
Department of CSE
Faculty of Science & Information Technology
Daffodil International University

\_\_\_\_\_\_\_\_\_\_\_\_\_\_\_\_\_\_\_\_
**(Name)**                                                                                                           **Internal Examiner**
**Designation**
Department of Computer Science and Engineering
Faculty of Science & Information Technology
Daffodil International University

\_\_\_\_\_\_\_\_\_\_\_\_\_\_\_\_\_\_\_\_
**(Name)**                                                                                                           **External Examiner**
**Designation**
Department of ----------------------
------------ University

©Daffodil International University                                                                                                       i

# DECLARATION

I hereby declare that, this thesis has been done by me under the supervision of **Professor Dr. Sheak Rashed Haider Noori, Professor and Head, Department of CSE** Daffodil International University. I also declare that neither this thesis nor any part of this thesis has been submitted elsewhere for award of any degree or diploma.

**Supervised by:**

___________________
**Professor Dr. Sheak Rashed Haider Noori**
Professor and Head
Department of CSE
Daffodil International University

**Co-Supervised by:**

___________________
**Dr. Abdus Sattar**
Associate Professor and Director
Department of CSE
Daffodil International University

**Submitted by:**

___________________
**Nafisha Tamanna Nice**
ID: 241-25-015
Department of CSE
Daffodil International University

# ACKNOWLEDGEMENT

Firstly, I want to express my thanks to the Almighty Allah for making me successful in the completion of my final year thesis.

I express my sincere thanks and great gratitude to **Professor Dr. Sheak Rashed Haider Noori**, **Professor and Head** of the Department of Computer Science and Engineering at Daffodil International University, Dhaka. He has a deep knowledge and a strong interest in the fields of "Big Data" and "Machine Learning" which is essential to the completion of this thesis. His untiring patience, academic mentorship, continual encouragement, diligent supervision, sound feedback, insightful counsel, thorough review of numerous inadequate versions at every stage have facilitated the completion of my thesis.

I want to mention my sincere thanks to **Dr. Sheak Rashed Haider Noori**, **Professor and Head** of the Department of CSE, for his invaluable help to complete my thesis and to the other academic members and staff of CSE department of Daffodil International University.

I need to extend my sincere thanks to my parents as it is the parent who is my best support and most dedicated to my work.

# ABSTRACT

Bug severity prediction is important in software maintenance, because it helps the development teams to prioritize bugs that have a significant impact on the operation, stability and security of the system. In large software projects bug repositories will grow at very rapid rate making classification of severity manual work labourious and unreliable and prone to human biasness. Many efforts have thus been dedicated on automated ways of severity prediction in the literature of software engineering research.This study compares different classifiers that are based on supervised machine learning algorithms for predicting bug severity levels using historical repository data from Eclipse Bugzilla. Evaluated methods range from linear classifiers, gradient boosting trees, distance method and transformer-based models, and text features, which are obtained from tokenization, TF-IDF, and n-grams and imbalance correction methods. Models were evaluated in terms of accuracy, precision, recall, F1 score, (AUC-ROC) and confusion matrix. Ensemble tree methods and DistilBERT achieved the top overall accuracy, while linear models performed best in recall of critical bugs which indicates some precision-recall tradeoff in imbalanced severity prediction. These findings provide useful actionable insight in choosing algorithms for automated bug triage, which can improve the quality of software through effective scalable prioritization.

# TABLE OF CONTENTS

# LIST OF FIGURES

# LIST OF TABLES

# CHAPTER 1

# INTRODUCTION

## 1.1 Introduction

In modern software engineering, predicting the severity of bugs has become very important as the level of severity of bug is generally used to determine the impact of bug on the functionality, performance, security, and overall user experience of the software. Traditional defect prediction attempts to determine whether a module is broken while severity prediction attempts to determine the severity of the defects that are reported. This helps software teams determine the priority debugging tasks they need to accomplish or the software resources they need to allocate and reduce the risks of the project.

There are sometimes thousands of reports in bug repositories for big projects, and it is hard to manually classify the severity of each one. To address this issue, supervised machine learning models have been created that can learn patterns from previous bug reports, and reliably make guesses about severity levels such as critical, major, minor or trivial. These models use text descriptions, information, and features of the project to make reliable and consistent predictions of severity.

Even if increasing number of people are interested in automated issue severity prediction, it is still difficult to select the best supervised issue severity prediction method. The behaviour of models like Logistic Regression, Linear Support Vector Machines (SVM), Passive-Aggressive classifiers, Stochastic Gradient Descent classifiers and Naive Bayes and non-tree methods such as k-nearest neighbors can vary greatly depending on the characteristics of the dataset, how features are incorporated, and data preprocessing. For example K-Nearest Neighbours vs tree-based ensemble methods like XGBoost, LightGBM,CatBoost deep learning based transformer-models such as DistilBERT. Comparison analysis is therefore necessary to establish whether supervised learning approaches achieved the optimal compromise of accuracy, robustness, and generalisation for software defect severity prediction problems.

## 1.2 Motivation

Software projects often get delayed and unsafe because of a poor bug prioritization. Understanding the severity of bugs is very important because bugs of high severity can affect the system's performance and user's trust of it. Manual classification has problems with inconsistency, time pressure, and experts.

Machine learning provides a solution that will automate the prediction of severity, guarantee speed, consistency and accuracy. This research aims to:

- Decrease the amount of time and effort required for manual triaging.
- Increase the accuracy of bug fixing prioritization during bug-fixing cycles.
- Improve general quality and reliability of the software.

This study aimed at comparing multiple supervised learning techniques in order to identify which ones are more effective, and therefore can be applied to real-world software engineering workflows.

## 1.3 Problem Statement

Software projects tend to be delayed and susceptible to security risks by poor bug prioritization. Although it is important to detect the bug, it is also important to determine the bug's severity since bugs with high severity could have a huge impact on the functionality of either system, or the user's trust. Manual bug severity classification is subject to inconsistencies and time pressure as well as differences in expertise. With the growing amount of bug reports, it is difficult to efficiently classify and prioritise bugs. Machine learning is providing an automated, accurate and consistent solution to predicting bug severity, allowing developers to prioritize bugs with the highest severity and enhancing software quality. This study is going to focus on tuning machine learning models for more accurate prediction of bug severity.

## 1.4 Research Objective

In this research, the goal was to develop a machine learning model for a predictive analysis of severity of software bugs. The idea is to be able to classify bugs by bug type and severity (using historical data of previously detected bugs as well as statistical metrics from the source code) so that developers can concentrate on critical issues in an early stage of the development process. This study aims to:

1. To identify the right machine learning algorithms for bugs severity prediction.
2. To measure the performance of these algorithms through different evaluation metrics such as accuracy, precision, recall and f1-score.
3. To establish predictive model to enhance the software testing and maintenance process by accurately predicting the bug severity.

These goals will allow research to improve the effectiveness and trustworthiness of software generation leading to more effective resource management and improving software quality.

## 1.5 Research Questions

This study has been driven by the following research questions:

1. RQ1: What ML algorithms are the better at predicting the severity of software bugs?
2. RQ2: How do we ensure the best possible precision and recall in high-severity bug-detection?
3. RQ3: What data preprocessing techniques can be applied to the algorithm to improve the accuracy of bug severity prediction?
4. RQ4: What are the difficulties in generalising the model to different software projects?

## 1.6 Report Layout

The report is divided into 6 chapters, and each of those addresses different aspects of my research:

Chapter 1 is an overview of the research topic, motivation, objectives, and research questions. Chapter 2 is a detailed review of work related to bug severity prediction, supervised learning techniques, text based classification, existing approaches in software analytics. Chapter 3 explains the research methodology, including the choice of the dataset, preprocessing steps, feature engineering, supervised learning models & evaluation procedures. The experimental results, model comparisons, performance analysis, and results discussion are presented in chapter 4. Chapter 5 discusses the societal, environmental, ethical, and aspects of sustainable of the research. Chapter 6 gives the conclusion, key insights, contributions, limitations & opportunities for future work.

# CHAPTER 2

# BACKGROUND / LITERATURE REVIEW

## 2.1 Introduction

The prediction of bug's severity is an important aspect of software quality assurance, given that it determines the way in which the time and financial resources are invested by development teams when maintenance tasks need to be addressed. General defect detection is detecting broken parts, whilst severity prediction is attempting to figure out how bad reported faults are in order to fix the important ones first.

Contemporary software projects generate a great number of problem reports by means of issue tracking systems such as Bugzilla, Jira and Mozilla repositories. These reports are generally unstructured, in plain English and hence it's difficult and not uniform to judge the urgency of these reports by hand. Judgement from individual to individual may differ depending on skillset and work load of developer as well as context of project. That can lead to slow responses to critical issues.

Automated methods to predict severity to overcome these problems have been given more and more attention, especially the ones based on supervised machine learning. By digesting such past problem reports, these algorithms are able to more quickly and reliably categorize incoming bug reports by severity. The recent improvements in NLP, as well as feature engineering, have made it even better at doing just that.

This chapter provides an overview of the overall research in the context of predicting the badness of software bugs, including the most common datasets and methods used for learning used in this domain, the techniques of preprocessing, in addition to the significant problems that motivated this study.

## 2.2 Related Works

Numerous research studies have been undertaken in order to better understand how to predict software bug severity. Here, we take a look to the various ways that have been used by experts to predict the severity of software defects.

A study in [1] applies Supervised Learning techniques to reopening bug report of software maintenance. Ten algorithms were tested including: kNN, SVM, Bayesian Network, Decision

Tree and ensemble algorithms like Bagging, AdaBoost and Random Forest. Among them Bagging and Decision Table (IDTM) got maximum F-measure 0.735 and 0.732 and surpassed the earlier approaches by 23.53%. The results emphasize that ensemble-based models, especially Bagging with traditional classifiers, can be very useful for improving the prediction accuracy. The authors conclude that machine learning can aid in the early detection of reopened bugs, which will lower the maintenance cost and increase the reliability of software, with future work to focus on optimizing parameters and other evaluation metrics.

In study paper [2], the authors discuss the nature and repair of bugs in big software systems. They talk about understanding bugs, sorting bugs and fixing bugs. The techniques that involve data mining, machine learning and natural language processing are discussed in this paper. It does this in the form of looking at 74 studies that were released in 2004. A bug report summary, bug duplicate detection, features prediction, bug developer suggestion techniques are all discussed. Bug fixing techniques such as IR based localisation and automatic patches generation are also covered. The study also finds things that have an impact on how well jobs are done and gives us ideas for how to continue studying deep learning in the future.

Aladdin Baarah et al. [3] used 1164 bug reports from JIRA of the INTIX Company. Eight classifiers were used for one-line bug summaries after preprocessing and TF-IDF features extraction and Information Gain feature selection. Bug severity was reduced to severe and non-severe classes. Logistic Model Trees had the best performance with 86.31% accuracy, and NBM had similar results. The study illustrates that automated severity prediction can be used to relieve manpower for manual triaging and enhance maintenance efficiency, with the suggestion of sentiment analysis for future improvement.

The authors of the study in [4] looks into how to use guided machine learning to find software bugs early on so that the quality of the software is better and the cost of making software is lower. With the knowledge about bugs and software measures in the past, Logistic Regression, Naive Bayes, Decision Tree and Random Forest were used, in addition to preprocessing, feature selection and class balancing using SMote. Random Forest got the highest accuracy of about 97% using 10-fold cross validation showing effectiveness of ensemble methods. The authors stress on feature engineering and recommend Neural Networks for future work.

Chaturvedi et al. [5] evaluates the machine learning techniques for problem severity prediction using textual summaries from both Open Source and commercial projects such as the National Aeronautics and Space Administration (NASA PITS), Eclipse, Mozilla and GNOME. Five classifiers were tested after pre-processing of the text and after feature selection. The results show that SVM has the highest accuracy, while the F-measure of Naive Bayes was better. Study shows that the performance is stabilised after a finite amount of features, which highlights the importance of careful preprocessing and classifier selection along with validation.

The authors in study paper [6] use machine learning models for predicting software bugs using datasets from the National Aeronautics and Space Administration (NASA). Using different evaluation criteria, Decision Tree and Random Forest algorithms reveals that they are better classifiers with almost 99 percent accuracy. The research highlights the importance of the software metrics, static code analysis, and effective evaluation techniques, and concludes that tree-based models are especially well-suited for fault prediction.

In the study paper [7] evaluate the results of Naive Bayes, Decision Tree, and Artificial Neural Networks for predicting software bugs using real world data sets. The results show that Decision Trees have a better overall performance and appropriate interpretability. The study has confirmed the viewpoints that the methodologies of machine learning exceed the traditional statistical approaches, and suggests further research into the use of ensemble methods and software measures.

The review in [8] takes a look at the problem of using supervised learning methods to predict the severity of bug reports based on the summaries of bug reports in the National Aeronautics and Space Administration's PITS database. We attempted to use a number of different classifiers using TF-IDF features and Information Gain selection. The results show that the performance of predictions stabilises at a certain number of features and that the predictions of binary or limited number of multi-class severity are better, whereas multi-class classification is more challenging because of the overlapping vocabulary.

Cetiner et al. [9] focus on improving the problem of class imbalance to predict software fault. Using SMote and PCA allows for more accurate predictions, mostly for Decision Tree and SVM models. The results demonstrate the importance of having balanced datasets and reducing the number of features.

In study paper [10] the concept of machine learning is adopted to make a guess about how bad the code smell is. Random Forest models are better than other classifiers especially when combined with SMote (for data balance) and Lime (for making sense of the results). The present study highlights the importance of the interpretability of models and balanced models in the setting of software quality evaluations.

Ha Manh Tran et al. [11] stated that the Random Forest model is a fantastic example since it works well with vast amounts of data and improves feature-based analysis, which is helpful for automatic bug management and software maintenance.

The authors [12] compare different machine learning classifiers that can forecast software bugs in open source Java projects. The results show that ensemble approaches, especially Bagged Trees, and SVM models are the best at a number of different measures. To make sure the results are reliable, statistical tests are applied.

The paper [13] addresses on the problems with classifying bugs by severity when the data isn't structured or is misclassified. The suggested Bug Classification and Reporting model gets higher precision, recall, and F-measure for a number of repositories. In future study, it may be possible to use deep learning to check its accuracy.

Both studies in [14] and [15] show that the Random Forest is the most similar to how a human expert would rate something, and that the quality of the datasets and the design of the features are critical for getting the right answer.

Akmel et al. [16] describe the results of an evaluation of ensemble-based defect prediction methods using NASA MDP datasets. The utilisation of ensemble learning for defect prediction is supported by the results, which demonstrate that the combination of AdaBoost and Decision Trees yields the maximum accuracy.

The survey in [17] shows that ML-based prediction aids in more efficient use of resources and more reliable software, and they corroborate the benefits and drawbacks of the various classifiers..

Meetesh Nevendra et al. [18] perform a study of deep learning methodologies in the area of software defect prediction. Despite the better accuracy of deep learning models over classic machine learning techniques, the issues such as class imbalance, feature complexity, and explainability remain as unsolved research topics.

The research in [19] proposes an ensemble-based study to enhance defect prediction by solving the problem of minority class bias. The proposed model shows good predictive performance, especially for AUC, which shows the importance of feature selection and the quality of the dataset.

The BPDET model introduced in [20] solves the class imbalance and overfitting issues for bug prediction. Assessed using the PROMISE datasets from the National Aeronautics and Space Administration, BPDET outperforms current models in a variety of evaluation parameters as well as showing a reduction in testing effort.

## 2.3 Scope of the problem

Bug severity prediction seeks to automatically assign the reported bugs into appropriate severity level, such that maintenance activities can be assigned accordingly. However, this task is difficult owing to various real world bugs report characteristics.

Some of the common types are that the structure of bug description will be informal, inconsistent in style, and it would not contain complete details so that relevant features cannot be extracted from these type of inputs. Severity labels are project specific issues of similar nature could have different severity levels. Moreover, the datasets that we use usually have an extremely strong class imbalance in that a critical bug might appear much less in comparison to a minor bug.

Thus, the scope of this problem is not only classified. This represents dealing with high dimensional unstructured text, imbalanced classes, meaningful feature selection, etc. Predicting models need to generalize to different software projects.

## 2.4 Challenges

While the use of machine learning can be very potential to predict severity of bugs, there are still many challenges to overcome:

**Data Imbalance:** Most of the data collected contain little number of high severity bugs, which could result in bias in models to predict lower severity levels.

**Feature Selection:** It is difficult to choose the most important textual and metadata traits because not all of the attributes assist in predicting severity in the same way.

**Model Generalization:** Differences in project structure, reporting methodologies and development procedures make it difficult to develop models that produce the same performance over multiple projects.

**Overfitting:** Good performance might have been seen for complex models on the training data but it may not be able to generalize well to unseen reports. To minimize this issue, there is a neccessity to use effective validation and regularization techniques.

Confronting these issues is important in order to develop reliable and scalable bug severity prediction systems.

# CHAPTER 3
# RESEARCH METHODOLOGY

## 3.1 Introduction

This chapter outlines the research methodology applied to employ machine learning techniques for predicting the severity of bugs of software systems. The intention is to use software metrics and historical bug data to make a good and effective model. The data collection procedure, machine learning methods used, data preparation procedures, and evaluation measures used to assess the model performance are all covered in detail in this chapter. In order to help developers prioritise important issues and improve the overall quality of the product, the goal is to offer a thorough method for the early estimation bug severity software development lifecycle.

## 3.2 Research Subject and Instrumentation

This study concentrates on employing machine learning techniques in predicting the severity of bugs in software systems. The goal is to create a model that uses software metrics and historical data and categorises defects into their different severity.

In order to do this, I analysed a dataset of 88,682 bug reports for Eclipse to look for some trends that could indicate high versus low severity of issues. Higher priority bug fixes is a big issue to the software industry, especially on big open-source software, like Eclipse, where developers are overburdened with manual triage. Teams can enhance response times and reduce downtime within the system if significant issues (blocking, critical, major) are prioritised when severity predictions are accurate. I have used machine learning techniques to predict severity probability from text descriptions and metadata given historical bug report data, with a focus on categorisation models. These models remove noise and detect significant correlations between bug attributes and bug severity results. The goal of this study is to evaluate the effectiveness of various machine learning approaches to the severity prediction and compare them with the transformer-based approaches to find the best methodology.

## 3.3 Proposed Methodology

In this research, I am proposing a model to predict the bug severity using machine learning algorithms. The idea is to categorize bugs by levels of severity according to the textual descriptions and different software measures.

I experimented with few machine learning techniques in order to make an efficient model for bug severity prediction. For reliable validation, K-fold cross validation (CV) for each machine learning method was used. The datasets underwent some preprocessing to handle missing data, encode categorical variables, and also SMOTE (Synthetic Minority Over-sampling Technique) for balancing the dataset. The data was then split into training and testing data for evaluation.

Several evaluation metrics have been employed including accuracy, precision, recall, F1-score and AUC. The model with the best performance in terms of these metrics was used for the final bug severity prediction.

### 3.3.1 Proposed Model Workflow

In order to make effective predictions of the software bug severity, a number of steps need to be followed. Below is Fig. 3.3.1, which shows the block diagram of the suggested model workflow to be used for this work:

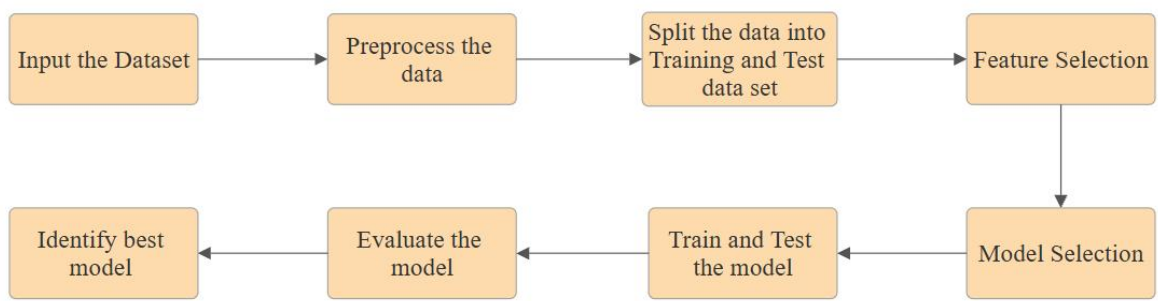

Fig. 3.3.1. Block diagram of the proposed model workflow

Below is Fig. 3.3.2, which shows the flowchart of the suggested model:

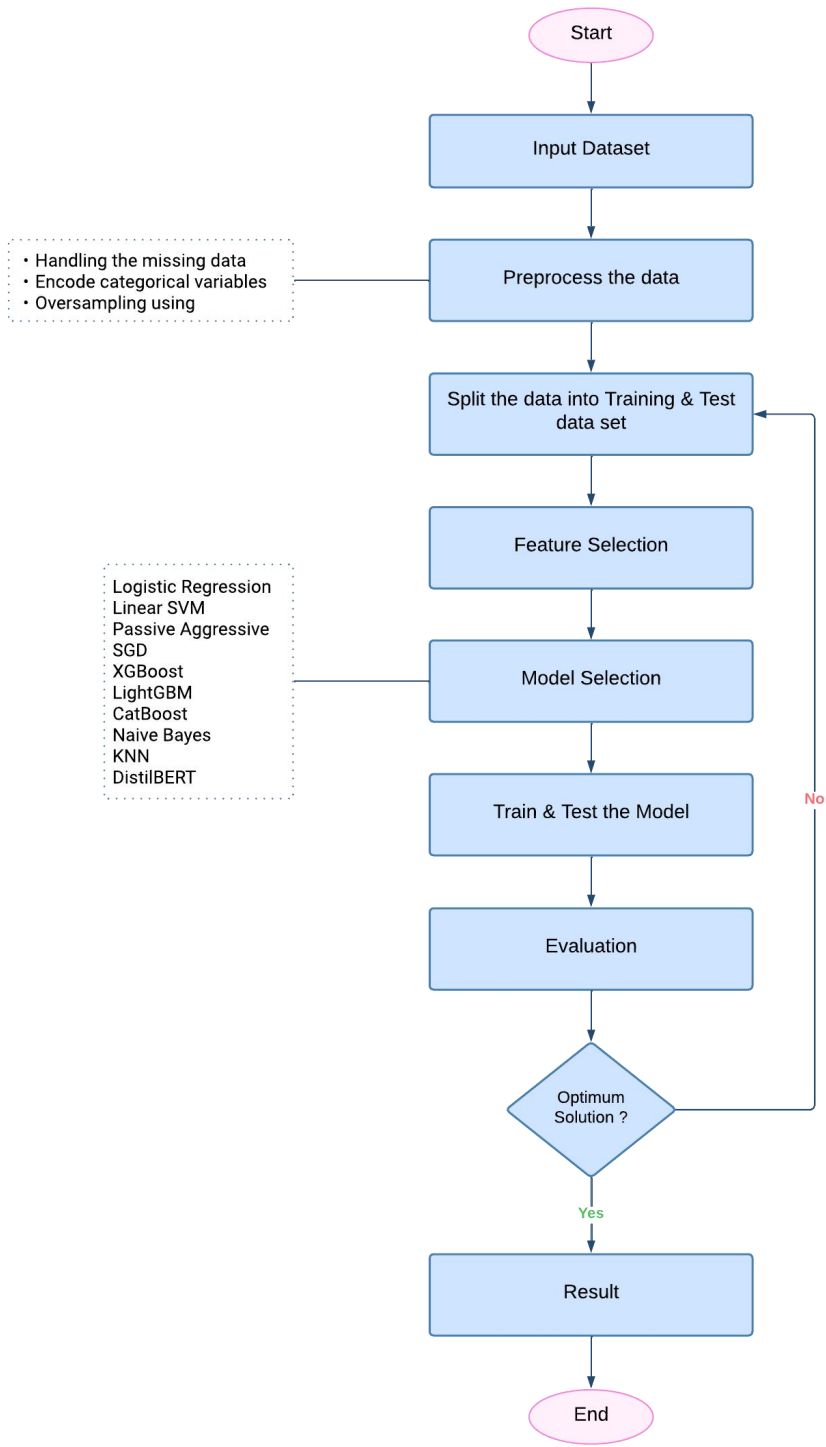

Fig. 3.3.2. Flow chart of the proposed model workflow

## 3.4 Data Collection Procedure

High-quality datasets are the first and foremost thing for an effective bug severity prediction models. The Eclipse Bugzilla dataset is a dataset with real-world bug reports with severity labels necessary for training classifiers. This dataset allows to analyze text descriptions and metadata in order to predict high vs low severity bugs.

**Dataset Source:**

Data is from the Bugzilla_Eclipse_Bug_Reports_Dataset.csv, accessed from Hugging Face. This platform contains machine learning datasets for NLP and classification. The dataset consists of 88,682 Eclipse bug reports with such fields as Project, Bug ID, Short Description, Bug Type, Resolution Status, Priority Label, and Severity Label.

### 3.4.1 Dataset

The Bugzilla_Eclipse_Bug_Reports_Dataset.csv gives a comprehensive real-world data for bug severity prediction. Hosted on Hugging Face, it consists of 88,682 reports of bugs for Eclipse from Bugzilla bugtracking system. There are structured metadata and text descriptions for each records for machine learning analyses.

The most important columns in the data set are as follows:
1. Project: The project of the bug.
2. Bug_ID: It is the unique identifier for the bug reports.
3. Resolution_Status: It is the current status of the bug.
4. Short_Description: Short description of the bug, which describes the bug issue.
5. Bug_Type: The type of the bug.
6. Priority_Label: It is the level of the bug's priority.
7. Severity_Label: It is the severity of the bug. This is the variable to be predicted by the prediction model.

The data set has a combination of both categorical and numerical features:
1. Categorical features such as "Bug_Type," "Priority_Label," and "Severity_Label."
2. Numerical features (not explicitly shown in the sample) May include time to fix or other features that may be derived or included as additional features during preprocessing.

The Severity_Label is the dependent variable in this research which will be used to form machine learning models for predicting bug severity based on other attributes. This dataset enables exploration of both text-based features (e.g. the bug description) and metadata (e.g. the bug type and priority) in order to build a comprehensive predictive model.

TABLE 3.1: ATTRIBUTE DESCRIPTION

| Attribute | Description | Data Type | Example Values |
|---|---|---|---|
| Project | The project associated with the bug (e.g., Bugzilla) | Categorical | Bugzilla |
| Bug_ID | Unique identifier for each bug | Numeric | 322082, 322412 |
| Resolution_Status | The status of the bug (e.g., FIXED, UNRESOLVED) | Categorical | FIXED |
| Short_Description | A brief description of the bug, providing details about the issue | Text | "Clean up user selection SQL", "Typo in error message" |
| Bug_Type | The type of the bug (e.g., Database, Documentation, Network) | Categorical | Database, Documentation |
| Priority_Label | The priority level of the bug (e.g., P1, P2, P3) | Categorical | P3, P4, P5 |
| Severity_Label | The severity of the bug (e.g., trivial, normal, critical) | Categorical | normal, trivial, critical |

## 3.5 Machine Learning Techniques

In this research, I have successfully forecasted the severity of software bugs utilising multiple machine learning approaches. Numerous types of learners are addressed in the methodologies presented such as: Tree-Based Learners, Ensemble Learners, Bayesian Learners, Neural Networks, Support Vector Machines, and Distance-Based Approaches. Out of these categories, I have selected 10 machine learning techniques that can accurately evaluate the severity of the software bugs. The algorithms used are shown in figure 1, with their respective categories. A list of each algorithm is found below.

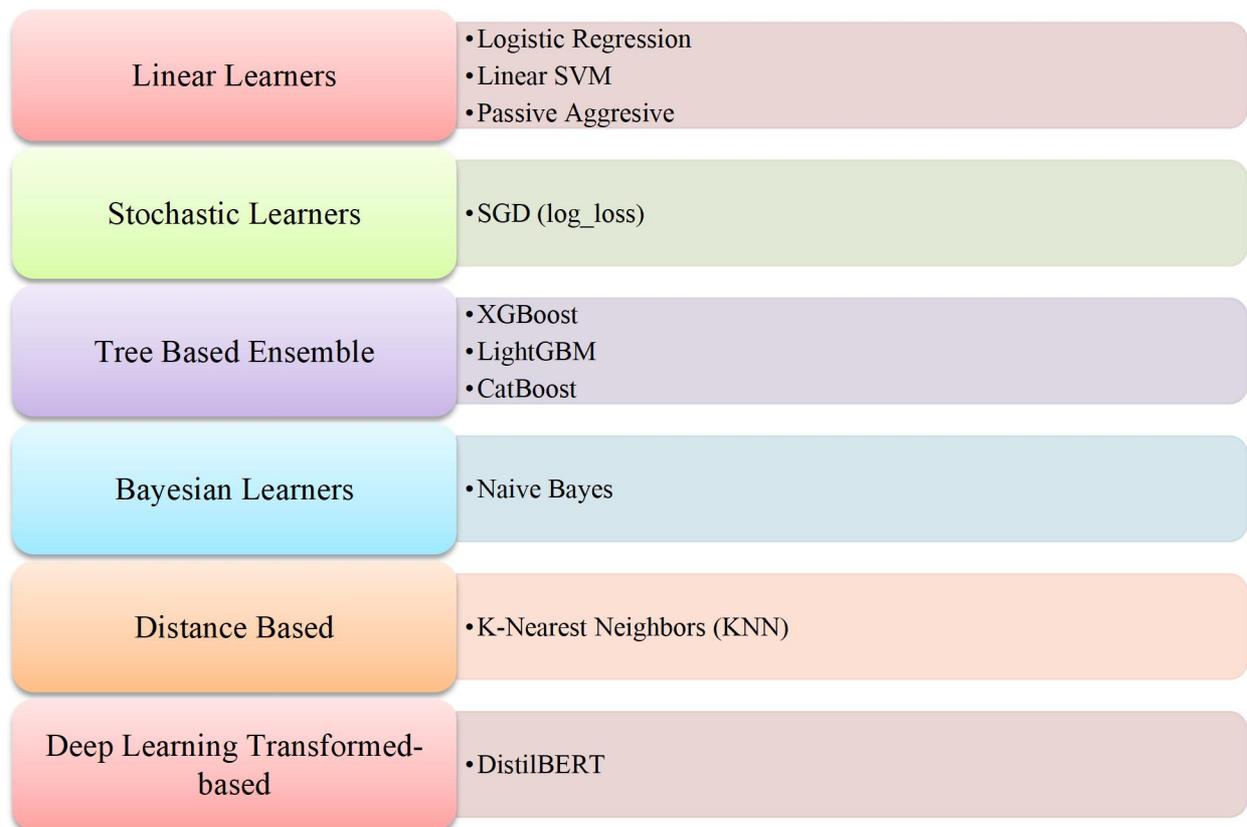

Fig. 3.5.1. Classification of ML Techniques

### 3.5.1 Linear Learners

**Logistic Regression**: This is a linear model that is used to place objects into two groups. It uses a linear combination of input features to make its guess as to how bad the bug might be. The model uses the logistic function to convert the weighted sum of the features into a number between 0 and 1 which indicates how likely it is that the bug is either very serious or not very serious.

**Linear SVM (Support Vector Machine)**: A linear SVM can be viewed as one type of classification technique where the hyperplane is found to best separate classes (in this case, the different degrees of problem severity). It works very well in areas with a lot of dimensions and CalibratedClassifierCV is used to fine-tune it to make better probability predictions. To separate bugs in groups according to their severity, the SVM algorithm uses a straight line as a decision boundary.

**Passive Aggressive Classifier**: This is basically an online learning system in which it adjusts its predictions based on the mistakes it makes. It is termed as 'passive' when there is no update when the model prediction is correct and 'aggressive' when an error occurs which leads to a significant update of the weights. This approach is efficient for large classification problems, and the use of class weights helps in solving class imbalance problems in bug severity prediction.

### 3.5.2 Stochastic Learners

**SGD (log_loss):** For binary classification problems, actually we use a SGD classifier with logarithmic loss. It is a better version of log loss function that computes the expected vs actual class probabilities. Stochastic Gradient Descent (SGD) It is a great approach to work with large dataset since it gets updated with small random sample of the data hence quick in training.

### 3.5.3 Tree-Based Ensemble

**XGBoost:** XGBoost is a framework of gradient boosting that makes trees one at a time. Each new tree corrects errors of the trees before. This ensemble method works on the idea of improving the model by training on the residuals (errors) of the previous trees. Random Forest is popular and widely used because of its good performance (high accuracy on test dataset), speed and scalability which is suitable for predicting bug severity over large datasets.

**LightGBM:** LightGBM, which stands for Light Gradient Boosting Machine, is another gradient boosting technique which was made to be quick and effective. The basic concept behind it is to use a method based on histograms that would help us accelerate the training process and especially when we have a lot of features and large datasets. LightGBM is a good technique for data with many dimensions, and it usually scales better, which makes it a good choice for predicting issue severity when the data is particularly complicated.

**CatBoost:** A gradient boosting algorithm that is optimized for categorical features which would be useful since bug tracking systems typically have categorical features (such as severity, status, etc) It directly deals with the categorical data, we have to do much less preprocessing. The CatBoost AlgorithmCatBoost is another tree-based algorithm which is great because of its robustness and ease of use, requiring little-to-no tuning to yield great results, making it effective at predicting bug severity.

### 3.5.4 Bayesian Learners

**Naive Bayes:** A probabilistic classifier called Naive Bayes is based on Bayes' theorem. This makes the calculation easier by assuming that the features are conditionally independent. But when it comes to predicting how bad a defect would be based on its observed properties (such its description, priority, etc.) measures the probability of each class (severity level). Naive Bayes is simple but very effective approach for the text classification like we are trying to do here: predicting the severity of bugs based on their description.

### 3.5.5 Distance-Based

**K-Nearest Neighbors (KNN):** KNN is a nonParametric learning algorithm, which learns from examples. KNN simply sorts the bugs according to the badness of their k closest neighbours in the feature space. This can be used to make a guess on how bad a bug is. The nearest neighbours by majority class give a bug a level of severity. This is a human-like way of looking at small data sets, in which the patterns are easy to see, because they are so simple.

### 3.5.6 Deep Learning Transformer-based

**DistilBERT:** A transformer-based model with a smaller size and faster training time as compared to BERT (Bidirectional Encoder Representations from Transformers). It has been enhanced for tasks such as text classification and in this case was applied to infer how bad bugs will be based on a written description. DistilBERT uses an improved model that can understand complex relationships between the data to better understand gun reports and predict how serious they are.

## 3.6 Data Preprocessing

For building reliable and accurate machine learning models efficient data preparation is important . This research included data pre-processing to make sure the data set was clean and balanced and appropriately prepared for predicting issue severity. The following steps were carried out:

### 3.6.1 Handling Missing Values

Missing values within the dataset could have an adverse effect on the effectiveness of machine learning models. The following were the strategies used for handling missing values:

- **Numerical Features:** For numerical attributes (e.g. features related to code complexity) missing values were replaced with mean of the respective feature. This is to ensure that the missing data does not affect the analysis or model training in a way that is biased towards the missing data.
- **Categorical Features:** For categorical attributes (e.g. "Severit_Label", "Bug_Type"), the missing values have been replaced by the mode (most prevalent value). This approach maintains the integrity of categorical variables and minimises the bias.

### 3.6.2 Feature Encoding

Many machine learning algorithms need numbers as input, hence category features were turned into numbers:

- **Label Encoding:** Categorical variables such as "Severity_Label", "Bug_Type" were label encoded in which each unique category of the variable would have an integer value assigned to it. This transformation was used on features that do not have an inherent order.
- **One-Hot Encoding:** Some features, such as "Project" and "Priority_Label" were one-hot encoded to create binary columns for each category. This technique does not use arbitrary numeric values for categorical features for which an inherent ordering does not exist.

### 3.6.3 Feature Scaling

To make sure that all the features contribute equally to the model, numerical features were standardised:

- **Standardization:** All the bug related features were standardised to have a mean of 0 and a standard deviation of 1. This prevents characteristics having bigger ranges overshadowing the model training process.

### 3.6.4 Text Preprocessing (For distilBERT)

Since DistilBERT was applied to text-based bug severity prediction, the following preprocessing steps were applied to the textual data:

- **Lowercasing:** "Short_Description" column textual entries were converted to lower case for maintaining consistency and avoiding case-sensitive conflicts.
- **Tokenization:** BERT tokenization was used for tokenizing textual descriptions into words or subwords. Here the texts is divided the text into tokens that can be processed by the model.

- **Padding:** To make sure that the transformer model always gets the same size input, all of the tokenised text was padded to a specified length, which was usually based on the length of the longest description in the dataset.

### 3.6.5 Feature Selection

Some features were considered as less relevant or redundant to predict bug severity. Features such as bug ID, resolution status, etc. have been discarded because they do not add much to the severity of a bug.

### 3.6.6 Splitting the Data

The data set was divided into the training and testing data sets to evaluate the model performance:

- **Training Set:** 80% data was taken for training the machine learning models.
- **Testing Set:** 20% of the data was set aside as a testing set that was used to test and evaluate the models after they were trained.

## 3.7 Statistical Analysis

This section contains a detailed examination of the statistics behind the machine learning models that were utilized to predict how bad an issue is. The primary task is to test the degree of applicability of each model, with a focus on the most significant assessment metrics. These metrics check on how well the models can guess how bad bugs are so that they can be trusted and work well.

### 3.7.1 Descriptive Statistics

Given that the dataset is mainly textual, no basic descriptive statistics for numerical and categorical features were computed in the traditional way (e.g. mean or median). Instead an emphasis was placed on the distribution of classes and textual data. However, for the models containing numerical features (e.g. in the case of distilBERT), these were processed as follows:

- **Textual Features:** The "ShortDescription" column which contains bug descriptions were tokenized and transformed using techniques such as TF-IDF (Term Frequency-Inverse Document Frequency) for textual analysis. Descriptive statistics like the most common terms, document frequency and feature sparsity were analyzed to determine the relevance of the text features.

- **Class Distribution:** For categorical features such as SeverityLabel and Bug_Type, a class distribution was checked to make sure that the dataset is not too imbalanced. The coloring of bug severity labels (e.g. "trivial," "normal," "critical") was important in terms of seeing how well the models would work with different classes.

### 3.7.2 Model Evaluation Metrics

Accuracy, Precision, Recall, F-measure is used for the testing of learning algorithms. The confusion matrix or the error matrix includes an overview of the results of the prediction of categorizations and used for the estimation of the effectiveness of every method. In multi output classification problems the evaluation of the model is very important.

There are True positive (TP), True negative (TN), False positive (FP) and False negative (FN) values in the confusion matrix.

- **True Positive (TP):** It is the predictions of the model that are same as what actually happened in the test.
- **False Negative (FN):** In this case, the model is wrong when it says that the result is negative when it is actually positive.
- **True Negative (TN):** This is when the model has predicted the estimated and test data as negative.
- **False Positive (FP):** The model is said to provide a positive result, where it should provide a negative one.

**Accuracy:** Calculates overall ratio of correct predictions of the model. It is calculated as:

$$\text{Accuracy} = \frac{TP+TN}{TP+TN+FP+FN} \quad (1)$$

**Precision:** It is defined as the ratio between the number of true positive predicted observations on the total number of positive predicted observations.

$$\text{Precision} = \frac{TP}{TP+FP} \quad (2)$$

**Recall (Sensitivity):** The number of positive examples that were correctly identified as positives with reference to the entire number of occurrences in real category.

$$\text{Recall} = \frac{TP}{TP+FN} \quad (3)$$

**F1 Score:** Harmonic The average of the precision and recall give you one number balancing the two.

F1 Score= $\frac{2 * Precision * Recall}{Precision + Recall}$ (4)

**AUC:** A measure of the ability of the model to separate different classes.

**Confusion Matrix:** A matrix which is a tabular version of the true positive, true negative, false positive and false negatives in a classification model.

### 3.7.3 Cross-Validation

We used the K fold cross validation method to test the models with k as 3. This method is followed for splitting the dataset into three smaller sets, or "folds." We only use one of the fold as a test set, and then we use the remaining 40% of the data to train the model. This approach ensures that the model is tested on a large number of different parts of the dataset, which makes the assessment outcome more reliable for determining how well a model functions. Cross-validation can help to prevent overfitting and ensure that the model is able to work well with new data.

### 3.7.4 Model Comparison

After analysing the models using the metrics above, the algorithms' performance was compared in the following ways:

**Bar Plots:** The accuracy and other evaluation metrics were visualized with the help of bar plots and compared for each of the models.

**Confusion Matrices:** We generated confusion matrices to display the true positive, true negative, false positive and false negative in each model.

By testing the models with these metrics, the best models to predict the bug severity were found.

## 3.8 Implementation Requirements

In order to implement machine learning for bug severity prediction, there are some hardware and software requirements that are essential for efficient model training, evaluation, and deployment. The following are the required specifications:

### 3.8.1 Hardware Requirements

- **Operating System**: Windows7 or Equivalent Operating Systems.
- **Memory**: Minimum of 4GBs of RAM, required to train the model.
- **Storage**: Minimum 300GB Free disk space.
- **Processor**: Multi core processor (multi core 2 +).
- **GPU**: It's optional (it's useful mostly for deep learning models, 2GB+VRAM).

### 3.8.2 Software Requirements

- **Python (3.x)** for implementation.
- **Libraries**: There are some of the libraries such as scikit-learn, xgboost, lightgbm, catboost, tensorflow, pytorch, pandas, numpy, smote (imblearn), transformers (for distilBert).
- **Development Tools**: Some of IDE needed such as VS Code or Jupyter Notebook or PyCharm for coding.

### 3.8.3 Internet Access Requirements

- **Dataset Repositories:** Internet access is also needed for downloading data set or other associated resources from public repositories like Hugging Face.

# CHAPTER 4

# EXPERIMENTAL RESULTS AND DISCUSSION

## 4.1 Introduction

This chapter discusses the results of the experiments that were done to see how well the machine learning models from Chapter 3 could predict how bad a defect was. It provides an in-depth view of how well each model performed, focussing on which criteria were used to determine how well the model did at predicting how severe a bug in software was. We trained and tested various algorithms on the data and then we were checking their results on the basis of accuracy, precision, recall, F1 score and area under the curve (AUC score).

## 4.2 Experimental Setup

The following steps are required:

1. **Data Preprocessing:** This was done like Fill Missing Values, Encoding of Categorical Variables, SMote used for balancing the data set.
2. **Training and Testing:** The models were trained using a 3- fold cross validation procedure, which ensured that the models were tested correctly. The data set was split into two sections. 80% of the data was used for training and 20% for testing.
3. **Model Selection:** Several machine learning models such as distilBERT, XGBoost, LightGBM, Logistic Regression, and Naive Bayes were selected with regards to their efficacy in bug severity prediction.
4. **Evaluation Metrics:** We have used accuracy, precision, recall, F1 score, AUC and Confusion matrix to rate the models.

## 4.3 Experimental Results & Analysis

This research experiment used 10 different machine learning models on the selected data set and predicted the severity of the software bug. We used a number of different measures to test these models, including accuracy, precision, recall, F1-score, AUC-ROC and confusion matrices. These measures were selected to ensure that we have a complete picture of how good each model is at predicting how bad an issue is.

TABLE 4.1: RESULTS FOR THE SELECTED DATASET

| Models \ Metrics | Logistic Regression | Linear SVM | Passive Aggressive | SGD | XG Boost | Light GBM | Cat Boost | Naive Bayes | KNN | Distil BERT |
|---|---|---|---|---|---|---|---|---|---|---|
| Accuracy | 85.07% | 89.62% | 84.60% | 85.79% | 90.41% | 84.48% | 90.17% | 89.94% | 89.94% | **90.48%** |
| Precision | 0.440 | 0.740 | 0.405 | 0.456 | 0.741 | 0.425 | 0.740 | **0.845** | 0.661 | 0.762 |
| Recall | **0.627** | 0.287 | 0.442 | 0.594 | 0.381 | 0.617 | 0.352 | 0.258 | 0.356 | 0.351 |
| F1 Score | **0.517** | 0.413 | 0.422 | 0.516 | 0.503 | 0.503 | 0.477 | 0.396 | 0.463 | 0.480 |

Table 4.1 summarises the performance figures of ten machine learning models, evaluated with the Eclipse Bugzilla bug severity dataset. DistilBERT with an accuracy of 90.48%, followed by XGBoost with an accuracy of 90.41% and CatBoost with an accuracy of 90.17% showed that transformer-based models and tree-based ensembles are very powerful in predicting bug severity.

Naive Bayes had the highest score of precision at 0.845 and thus showed its ability to produce the least number of false positives for high severity issues.

Logistic Regression showed the score of 0.627, showing its effectiveness in identifying high severity problems. Furthermore, Logistic Regression achieved the highest F1-score (0.517), which provides a good balance between precision and recall.

These results show the power of some models in bug severity prediction, with DistilBERT and XGBoost showing high accuracy, although Logistic Regression is best for predicting high severity bugs. Models like Naive Bayes show better precision, while Logistic Regression achieves the right balance between a number of requirements.

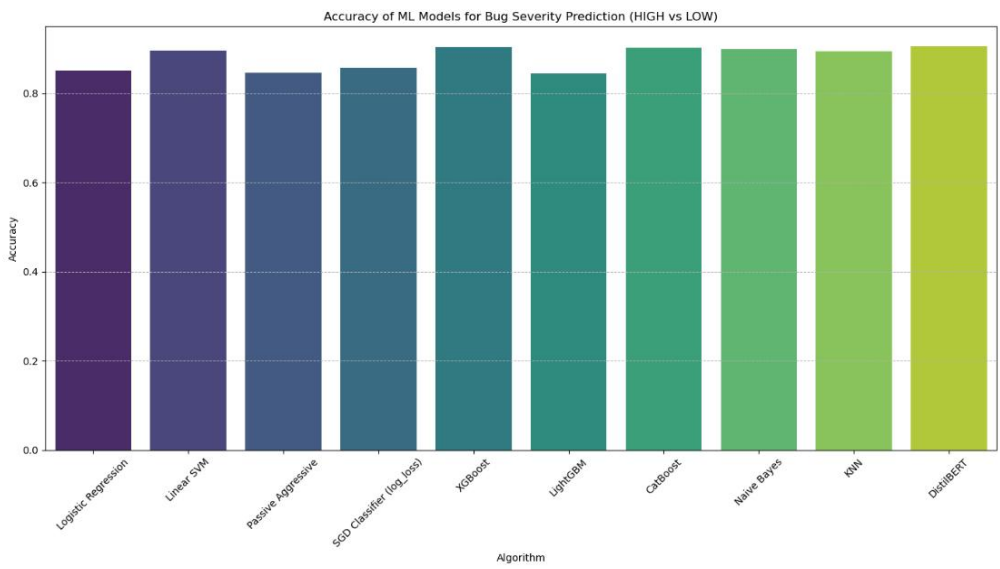

Fig. 4.3.1. Accuracies of ML Techniques for Software Bugs Severity Prediction

We utilised confusion matrices to show how well the models performed by showing how many true positives, false positives, true negatives, and false negatives there were.

These matrices provide a full analysis of the efficacy of each of the models to classify high-severity and low-severity bugs. They allow us to assess the ability of each model to correctly identify high-severity bugs, which is crucial for bug prioritisation and allowing successful triage in software development. The following matrices are used to show the predictive performance of each model over the entire dataset.

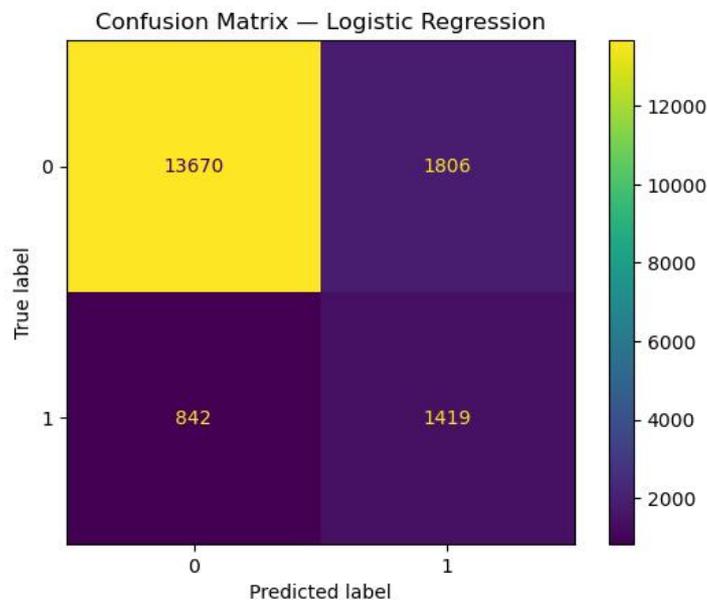

Fig. 4.3.2. Confusion Matrix for Logistic Regression

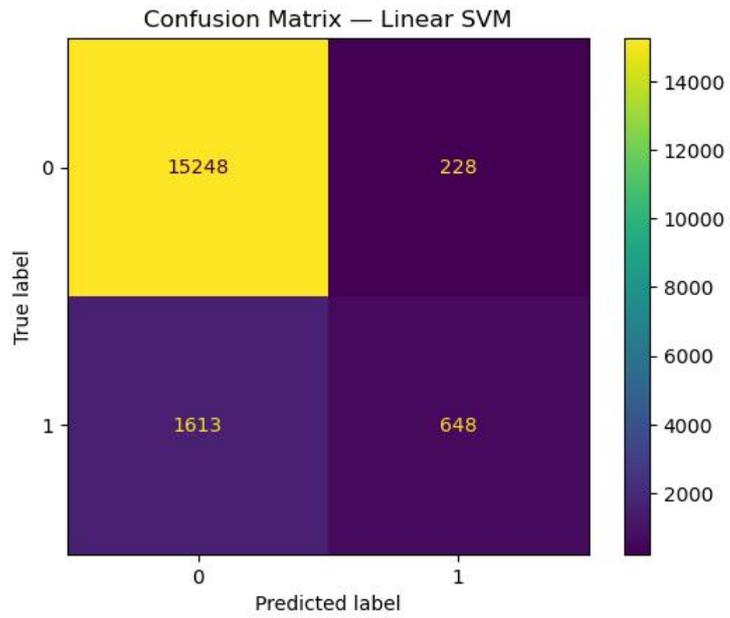
Fig. 4.3.3. Confusion Matrix for Linear SVM

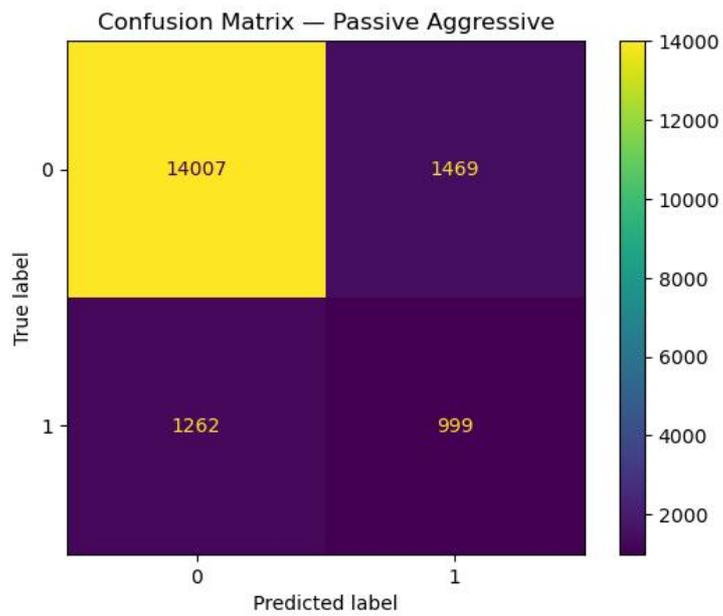
Fig. 4.3.4. Confusion Matrix for Passive Aggressive

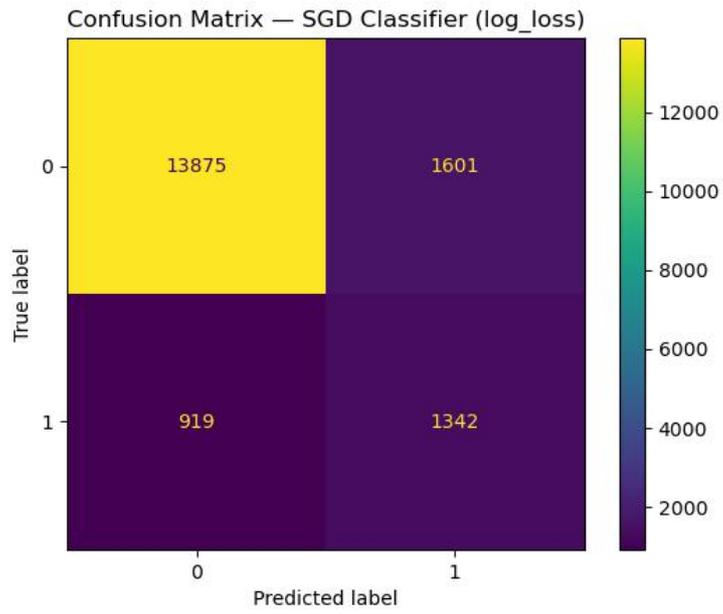
Fig. 4.3.5. Confusion Matrix for SGD

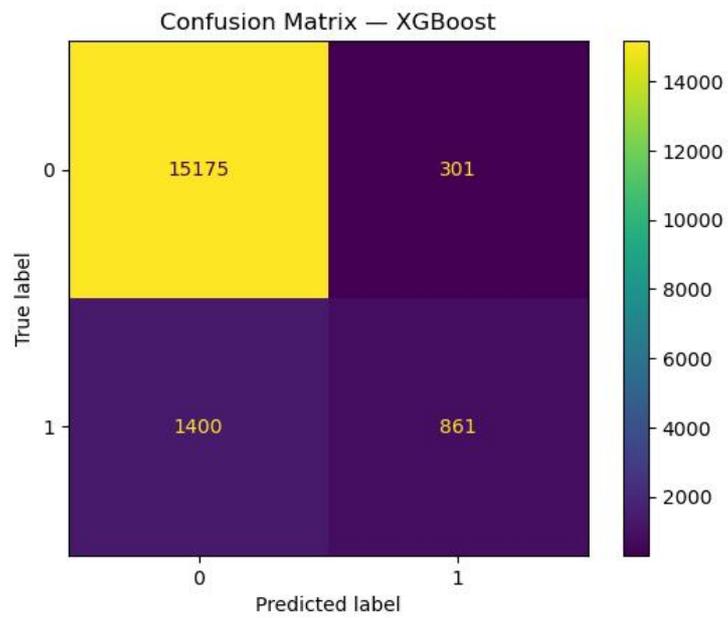
Fig. 4.3.6. Confusion Matrix for XGBoost

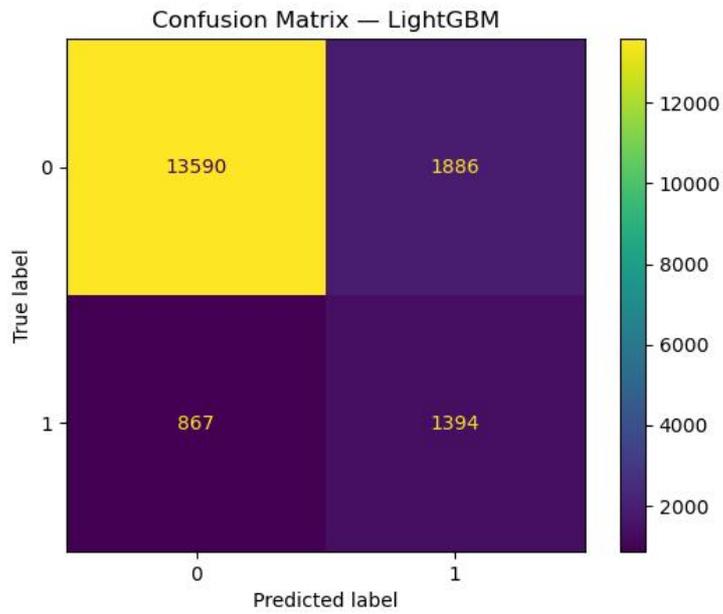
Fig. 4.3.7. Confusion Matrix for LightGBM

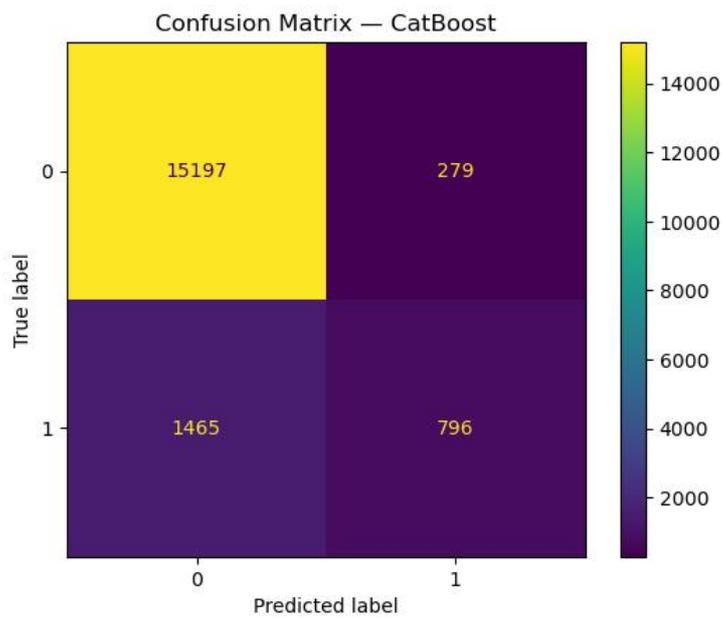
Fig. 4.3.8. Confusion Matrix for CatBoost

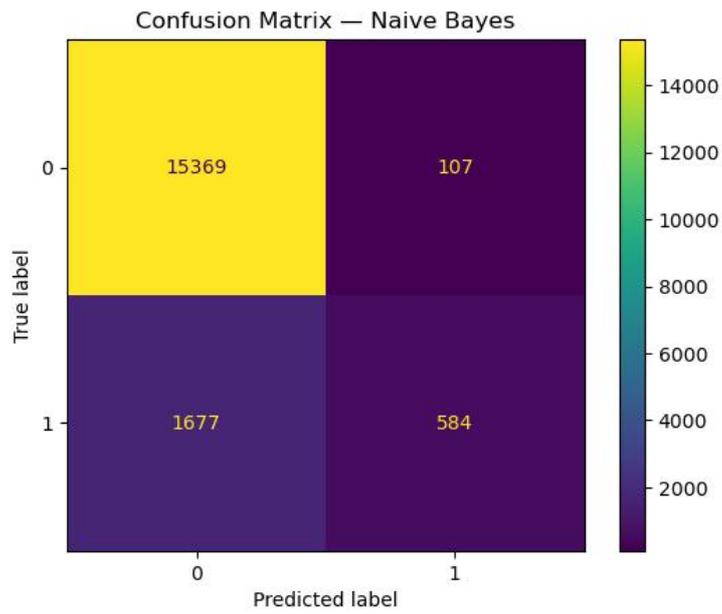
Fig. 4.3.9. Confusion Matrix for Naive Bayes

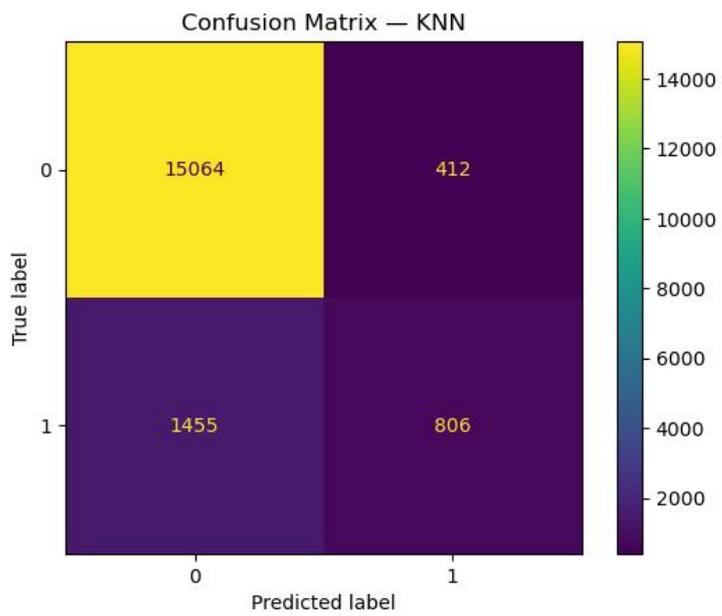
Fig. 4.3.10. Confusion Matrix for KNN

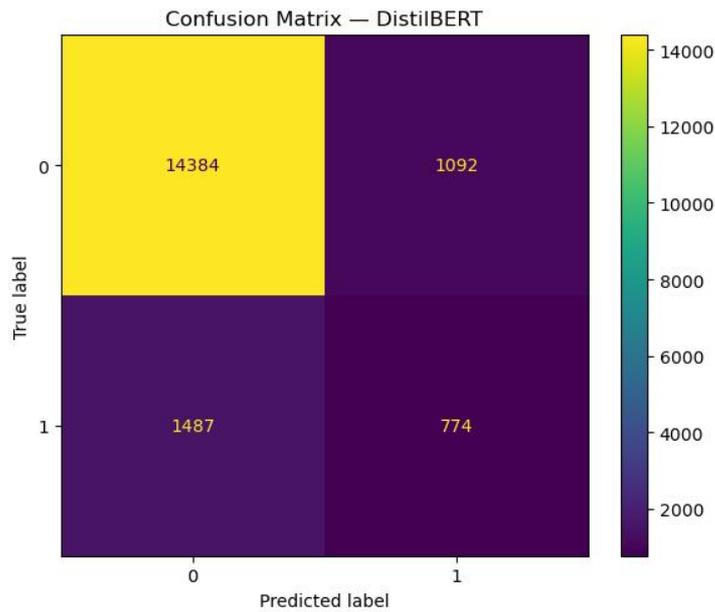

Fig. 4.3.11. Confusion Matrix for DistilBERT

In this study 10 machine learning models were tested in order to predict the severity of software bugs. In order to understand how each model is able to generalize on unseen data, we compared training accuracy validation accuracy with various training size. The following analysis summarizes trends that are observed across all models:

- **Logistic Regression:**

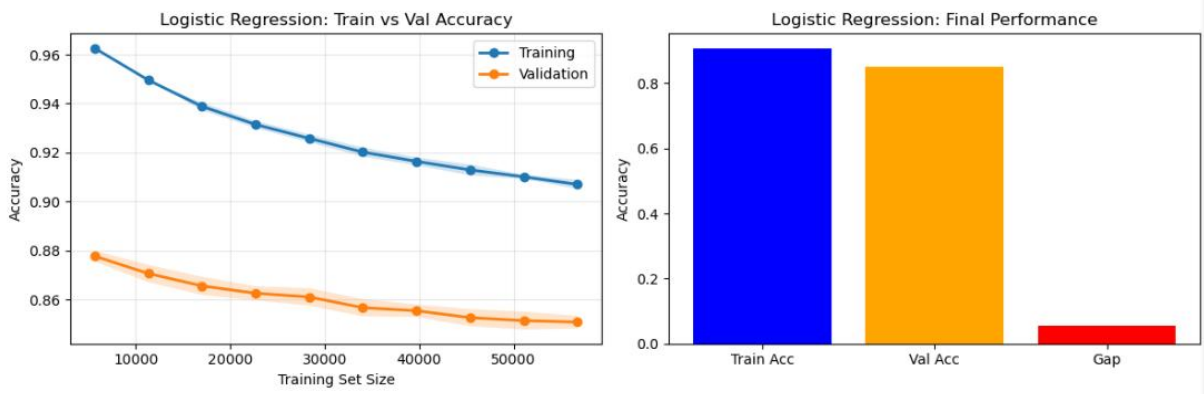

Fig. 4.3.12. Training vs Validation Accuracy for Logistic Regression

While there is higher training accuracy in this model, the model works well for recall to predict high severity bugs. The accuracy of the validation is a bit lower but that means that the model is focused on detecting critical bugs which is needed to ensure that you minimize false negatives.

- **Linear SVM:**

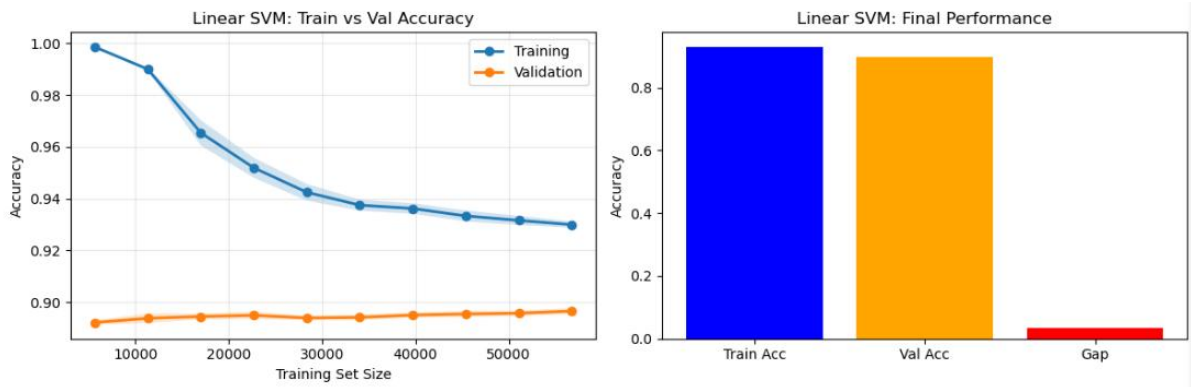

Fig. 4.3.13. Training vs Validation Accuracy for Linear SVM

Linear SVM is also showing good training accuracy. The model is also consistently better in smaller datasets although validation accuracy is reduced, it retains a good balance between precision and recall which makes it suitable for scenarios requiring both types of metrics to be important.

- **Passive Aggressive:**

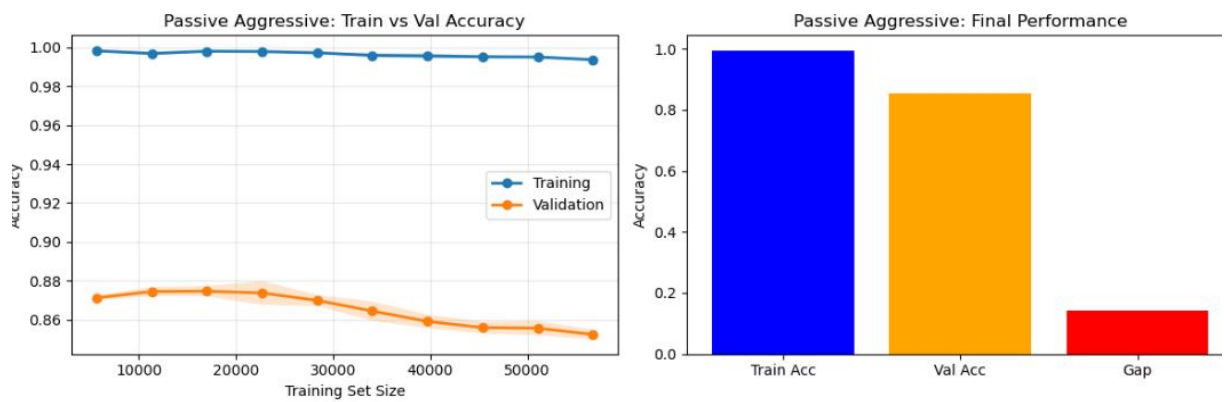

Fig. 4.3.14. Training vs Validation Accuracy for Passive Aggressive

This model has good performance, the training accuracy is high. The validation accuracy gets low as the training set size increases, but it is a good option to go for if you want to get fast, real-time predictions especially with large-scale data.

● **SGD Classifier:**

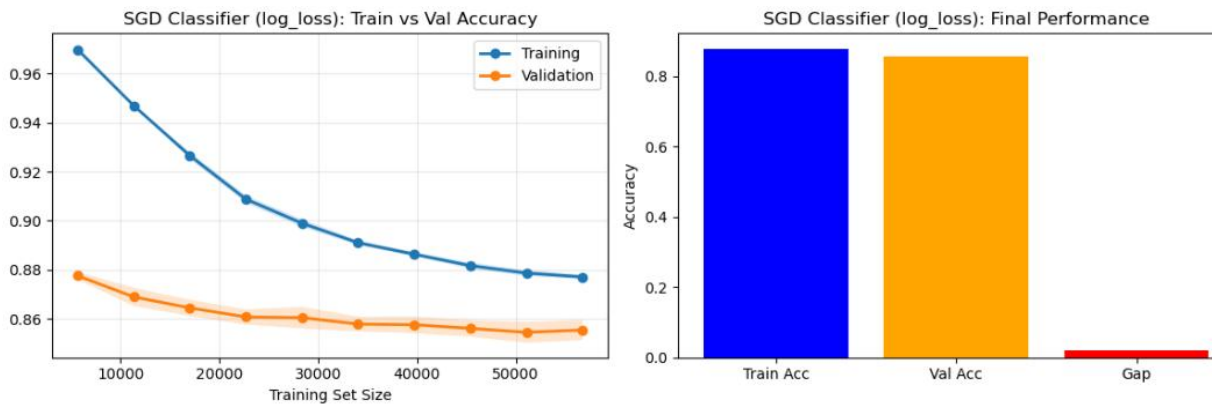

Fig. 4.3.15. Training vs Validation Accuracy for SGD Classifier

The SGD Classifier is efficient and has good training accuracy with competitive validation accuracy. Its speed and efficiency in handling large data sets make it an ideal choice for applications that demand fast predictions, in particular: Real time bug tracking systems.

● **XGBoost:**

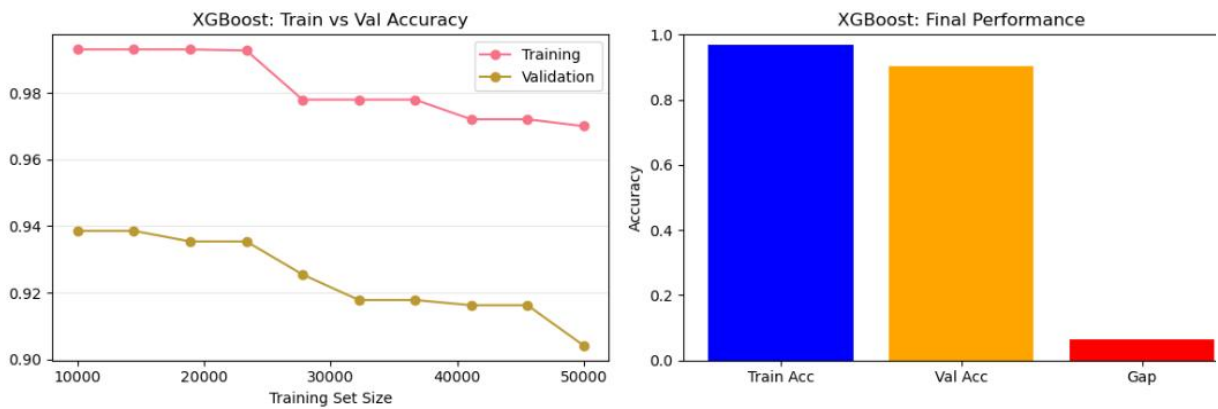

Fig. 4.3.16. Training vs Validation Accuracy for XGBoost

XGBoost is the highlight of the accuracy and the precision is very high. The model works well on both the training and validation data, so it has a good generalization. It is a top choice to ensure maximum accuracy and precision in bug severity prediction and reliable bug prioritization.

- **LightGBM:**

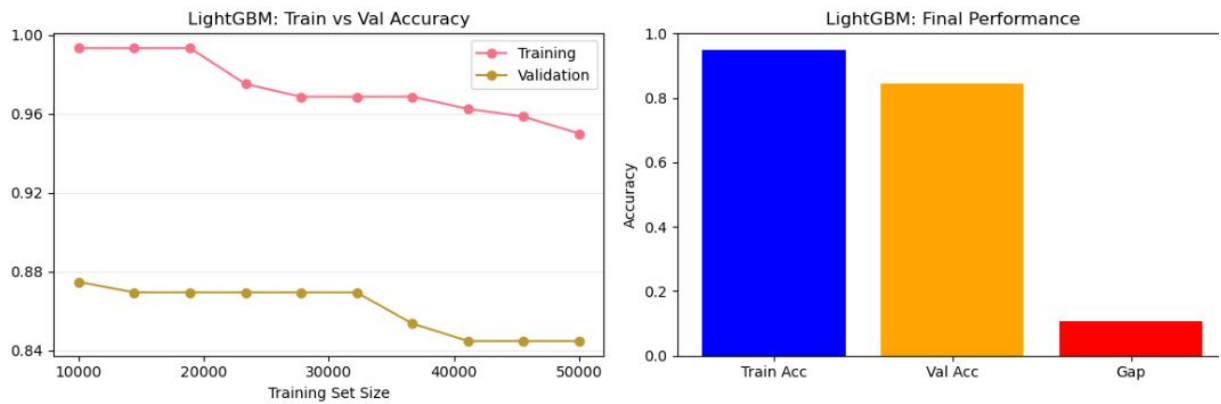

Fig. 4.3.17. Training vs Validation Accuracy for LightGBM

The model has superior performance and stable training and validation accuracy. The model shows considerable potential in dealing with huge datasets, and the model is capable of handling high-dimensional data, making it an outstanding choice for predicting bug severity of various projects.

- **CatBoost:**

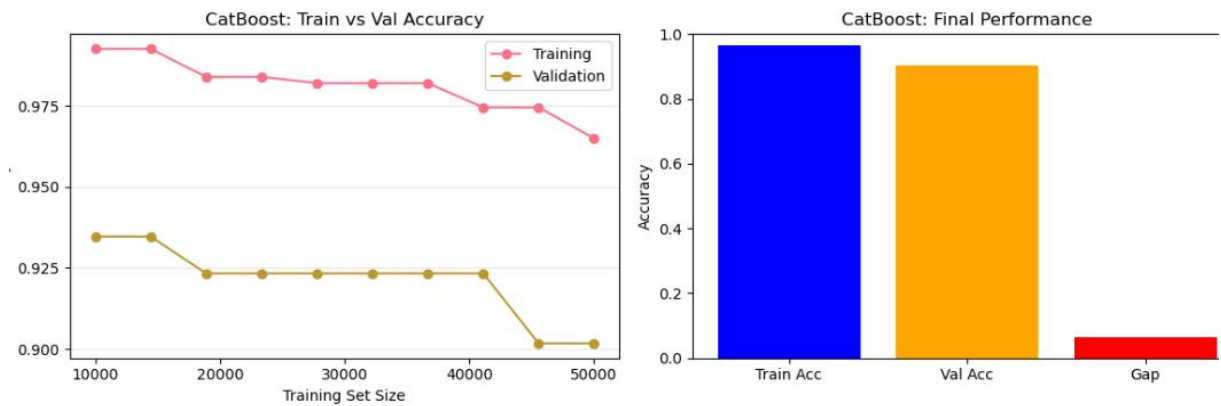

Fig. 4.3.18. Training vs Validation Accuracy for CatBoost

CatBoost has impressive performance with little training-validation gap. This model is extremely good with categorical data and is a good choice for predicting bug severity in software system that has different types of data and different degree of complexity.

- **Naive Bayes:**

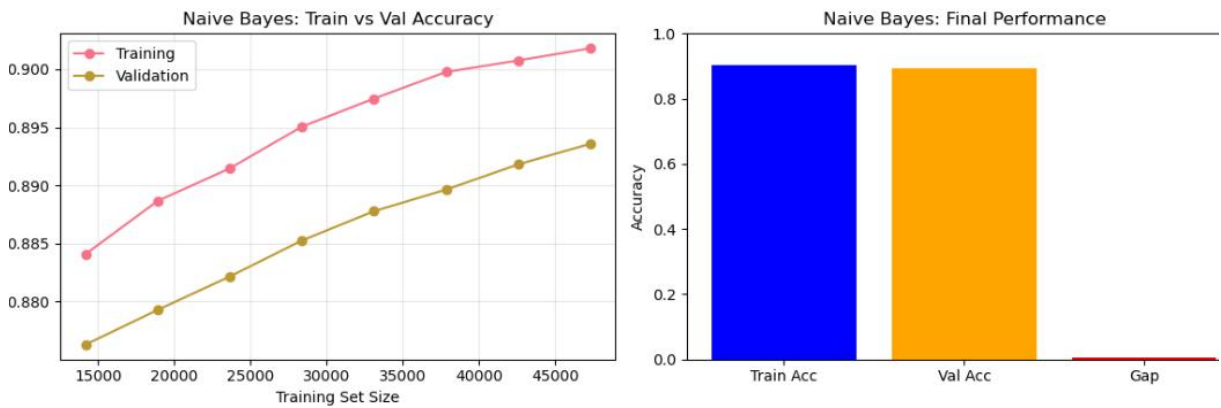

Fig. 4.3.19. Training vs Validation Accuracy for Naive Bayes

The model has good generalization capability with consistent train and validation accuracy. While its overall performance may not be as high as other models, its simplicity and effectiveness in classifying text make it a reliable choice for predicting bugs that are not as complex or when the interpretability of the model is an important factor.

- **KNN:**

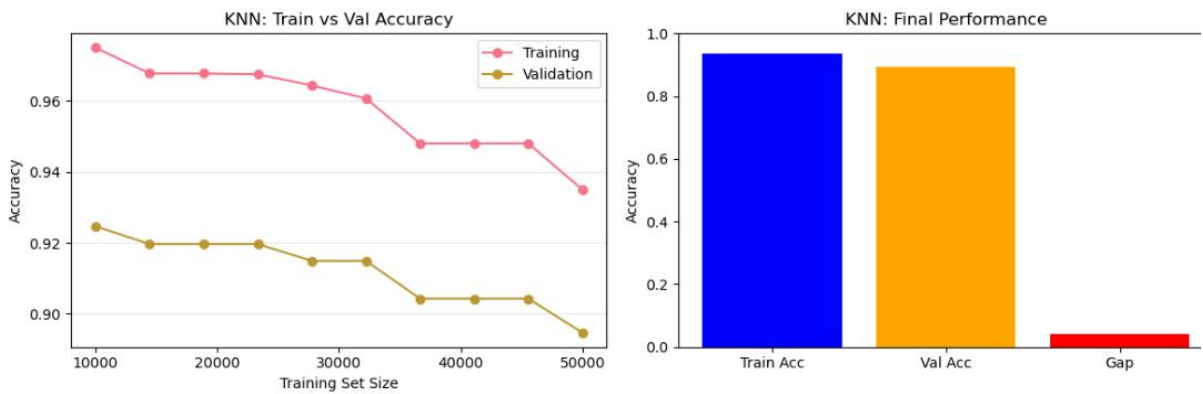

Fig. 4.3.20. Training vs Validation Accuracy for KNN

KNN is also consistent in the performance of the training set and the validation set. While it has smaller gap as compared to more complex models, it's a great choice for simplicity and interpretability giving fast predictions to smaller data sets or situations where you need model interpretability.

● **DistilBERT:**

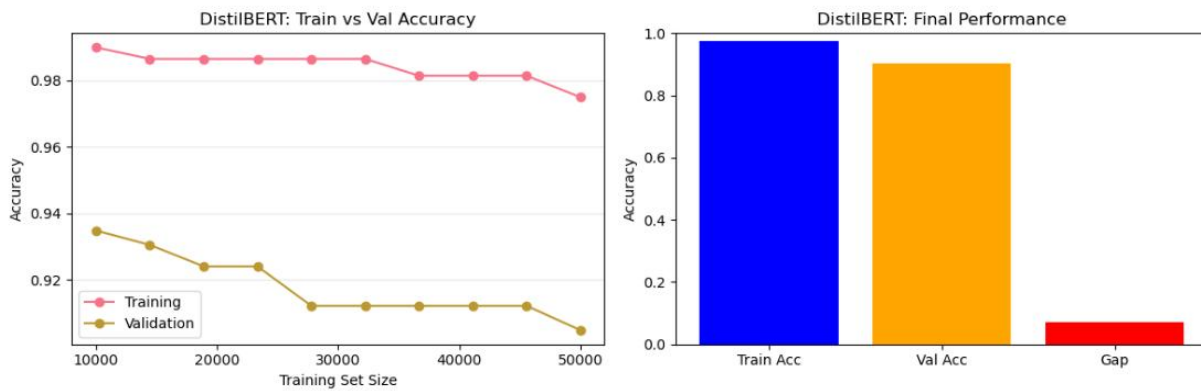

Fig. 4.3.21. Training vs Validation Accuracy for DistilBERT

DistilBERT achieves better performance, as it has a higher accuracy and precision. The model handles complex features of text and has a great balance between training and validation accuracy. It shows a high potential for text-based bug severity prediction, making it the most efficient model for high severity bug prediction.

Along with accuracy, precision, recall and the F1 score, confusion matrices are also included for each model to provide a more comprehensive view of how well the model works. We can use the confusion matrices to plot the numbers of true positives, false positives, true negatives and false negatives. This allows us to determine how well each model is able to predict high-severity and low-severity bugs.

## 4.4 Discussion

The best classification results for each evaluation metric are highlighted in bold in Table 4.1, which summarizes the performance of ten supervised learning algorithms on the Eclipse Bugzilla bug severity dataset. All models were trained using TF-IDF features derived from the brief bug descriptions and types, and assessed on a standardised test split to ensure equal comparison among algorithms.

● **Model Performance and Effectiveness (RQ1)**

In terms of overall accuracy, DistilBERT (90.48%) and XGBoost (90.41%) had the highest scores, suggesting that transformer-based models and gradient-boosted tree models are the most effective in general bug severity prediction in this study. These models performed better than the rest in terms of classification accuracy, which indicates that the model distilBERT with its ability to comprehend complex textual data, and XGBoost with its ensemble learning technique are suitable for bug severity prediction in a real world scenario.

CatBoost (90.17%), Naive Bayes (89.94%), KNN (89.94%) and Linear SVM (89.62%) also demonstrated a good accuracy, but were slightly less accurate than DistilBERT and XGBoost. While they still gave good predictions, the small performance decrease shows that tree-based models (CatBoost), Naive Bayes, may not handle textual features to the same degree of complexity as transformer-based models such as DistilBERT.

- **Preprocessing Techniques (RQ2)**

The findings highlight the importance of organised preprocessing to make machine learning models more effective.

The combination of text cleaning, TF-IDF vectorisation and the use of bi-grams significantly improved the accuracy and overall effectiveness of the models. These preprocessing strategies helped the models to better understand the textual descriptions of the defects, which led to better models in terms of efficacy in determining the severity of the defects.

It was very important to use SMote, Synthetic Minority Over-Sampling Technique to fix the class imbalance problem in the dataset. This is because only 12.8% of the cases had problems with categorisation of severity to "HIGH". SMOTE helps to make synthetic samples for the minority class. This helps ensure that the models have a chance to see a sufficient number of the HIGH-severity flaws, and don't take to favouring the low-severity issues.

Adding class weighting in a balanced way in models such as Logistic Regression and SGD Classifier helped in making the model work a lot better on the unbalanced dataset. This allowed these models to work on correctly classifying the minority class (high-severity bugs).

- **Precision and Recall for HIGH-Severity Class (RQ3)**

Going into precision and recall, for the HIGH-severity class, the obtained results offer interesting insights about the model performance:

Naive Bayes showed the best precision (0.845), indicating that it is the most conservative of the model and generates the least false positives for bugs of the HIGH-severity category. This is an important aspect for situations where minimizing the risk of misclassifying low severity bugs as high severity is important.

Logistic Regression, however, had a better performance in recall (0.627) and F1-score (0.517) meaning that it was better at correctly identifying the bugs that have the severity of the HIGH class, at the cost of some false positives. This ability to find a greater number of high-severity bugs gives Logistic Regression an edge in situations where it is important that a number of high-severity issues be found, such as in critical bug triage and prioritisation.

- **Generalization Across Projects (RQ4)**

One of the issues revealed in this research was generalizing the model for other software projects. Models like Logistic Regression and Naive Bayes were excellent in accuracy and bad on recall while the models like distilBERT and XGBoost were excellent on accuracy but found it difficult on recall. This means that some of the models may be better in terms of overall accuracy, but may not generalize very well across different types of software projects or bug distributions.

For bug detection in high severity cases, models such as Logistic Regression are a very good fit where recall is a priority and precision can be given a back seat. However, for the cases when overall accuracy and precision is more important one can go for the models like XGBoost, CatBoost and DistilBERT which is a good balance of both.

The results suggest that in practical applications model selection should be determined on a project specific priority. For projects aiming to reduce false negatives (missing high-severity bugs) it would be most appropriate to use models such as Logistic Regression or SGD. For projects where overall bug severity prediction accuracy and precision are important, distilBERT, XGBoost or CatBoost would be preferred.

©Daffodil International University                                                                                                            46

# CHAPTER 5

# IMPACT ON SOCIETY, ENVIRONMENT, AND SUSTAINABILITY

## 5.1 Impact on Society

The fact that a software bug severity can be accurately predicted can have a profound effect on software development methodologies. By focusing on the high severity issues first in the development cycle, developers can focus on resources to fix key issues, which reduces any downtime and improves the overall quality of software products. The results of this study can:

- **Increase Software Quality:** By identifying and fixing bugs with high severity at an early stage, developers can ensure more reliable, secure and stable software. This is especially important in mission-critical applications, such as healthcare, finance and telecommunications.

- **Enhance Development Efficiency:** Developers can save time working on issues that are not critical to development, leading to faster development cycles. This helps in meeting deadlines, and helps to improve time-to-market of new software releases.

- **Improve User Experience:** Responding to high severity bugs in a timely manner will result in a better user experience, with fewer crashes, bugs, and disruptions. This is especially true for end-users who rely on smooth and efficient software for their daily activities.

- **Aid in Bug Prioritization:** Utilising machine learning to predict the severity of defects helps the organisation to create more efficient bug triaging procedures, giving more importance to the resolution of the critical issues and thereby, facilitating effective allocation of resources.

## 5.2 Impact on Environment

Although immediately the bugs severity prediction may not appear to have that much impact on the environment, it may contribute to sustainability indirectly in a variety of ways.:

- **Resource Optimization:** With proper prediction and prioritization of bugs, software teams can limit the resources they allocate to the fixing of less important bugs. This makes for less computational waste and a more efficient upkeep of the software, which is a contributing element in more efficient energy and hardware utilization.

- **Saving time:** With better predictions, fewer and more critical debugging efforts are concentrated on the critical issues, saving time spent in whole bug fixing. This can lead to the indirect use of less energy as it will prevent having to run multiple versions of software in order to test different fixes.
- **Sustainable Software Practices:** Efficient software development practices that prioritizes high severity bugs can contribute to the development of more sustainable technologies that reduces the environmental impact an ongoing software maintenance, as well as reducing the energy cost of development environments.

## 5.3 Ethical Aspects

The use of machine learning techniques to predict software defects raises a number of ethical considerations:

- **Bias and Fairness:** Machine learning models, for example, those used for bug severity prediction, can even accidentally perpetuate biases that exist in the data they are trained on. If the data set includes a skewed set of bug reports, then the model may be biased towards some categories of defects to others. It is vital to make sure that the model does not generate unfair biases that can damage specific users or software endeavors.
- **Transparency:** It is necessary to ensure that the ML models used for bug severity prediction are transparent and that their decision-making processes are understandable. This will help developers and stakeholders to put their trust into the predictions and make well-informed decisions based on them.
- **Data Privacy:** The data set used to perform this study is a set of bug reports, which at times may be associated with confidential information. It is important to manage and anonymise such data in the proper way to comply with data privacy regulations.
- **Accountability:** While there is no denying that machine learning models could have a role to play in bug severity prediction, there should always be a human in the loop when it comes to final determinations. As model developers, they should be responsible for validating the predictions and ensuring that the models are used ethically and responsibly.

## 5.4 Sustainability Plan

To make sure that this research has a long-lasting impact and can continue, there are several steps that may be taken to make sure that the models created are kept and improved:

- **Continuous Learning:** Machine learning models, such as the ones used in this study, should be continuously trained on new data in order to adapt to changing software environments. By adding feedback loops and retraining the models with new data, the models can keep their relevance and accuracy over time.
- **Open-Source Development:** By releasing the models and datasets to the public, developers of various backgrounds can help to develop the models for improvement. This will create a sense of collaboration and ensure that models are not only available but continuously improved.
- **Collaboration with Industry:** Working with the software development industry to test and implement these models in real-world settings will help to validate the findings and ensure that the models are attending to industry needs. Partnerships with companies could guarantee the extensive use of machine learning for bug severity forecasting further adding to the efficiency and sustainability of software development.
- **Monitoring and Evaluation:** Regularly monitoring and evaluating the performance of the models in the real world will ensure that they are effective. Feedback from users and developers should be included to fine-tune and improve the models to ensure they are relevant and impactful.

# CHAPTER 6

# CONCLUSION AND FUTURE WORK

## 6.1 Summary of the study

This study mainly focused on developing and evaluating machine learning models that will be used to predict the severity of software bugs. I used the bug reports dataset of Eclipse Bugzilla to try out some machine learning approaches to predict bug severity. I then used a number of metrics, such as accuracy, precision, recall, F1-score and AUC-ROC to see how well these methods worked.

The research shows that the transformer-based (e.g. DistilBERT) and tree-based ensemble (e.g. XGBoost, CatBoost) models provide better accuracy and performance making them extremely useful in predicting bug severity. Logistic Regression was the best model for identifying high-severity bugs, with a great recall and F1-score, showing that it is important to be able to recognize important matters as soon as possible.

The research shows that effective data preprocessing steps, such as handling class imbalance in the data using SMOTe and TF-IDF for text vectorisation, are necessary for improving the performance of the model.

## 6.2 Conclusions

The research conducted has the following conclusions:
- **Model Effectiveness:** DistilBERT & XGBoost are the best models to predict bug severity with the highest overall accuracy. They work well with text data and also work well with complex bug descriptions.
- **Best Models for High-Severity Bug Detection:** Logistic Regression proved to be the most effective model for the detection of high-severity bugs as it scored the highest recall and F1-score.
- **Preprocessing Techniques:** Data preprocessing techniques such as SMOTE to balance the data set and TF-IDF to vectorize the text data helped the model to generalize and perform well on different levels of severity.

- **Challenges in Generalization:** While the models performed well on the dataset, there is a possibility that their performance may differ when applied to other software projects with different characteristics. Therefore, the generalization across projects needs careful model selection depending on the project's priorities (e.g. accuracy vs. recall).

## 6.3 Implication for further study

This research has been useful for predicting how bad problems will be, however there are several parts of it that may be looked into more in the future:

- **Expanding Datasets:** Future studies will have to consider a broader range of datasets from different software workplaces in order to evaluate the generalisability of the models in different areas. We can better determine the strength of these models using datasets from other fields.
- **Advanced Preprocessing Techniques:** Further research can be done on how to better preprocess text, by using word embeddings (e.g. Word2Vec, GloVe) or more advanced transformers (e.g. BERT). These could make the model more aware of the contextual relationship in bug descriptions and, as a result, make the model more capable on bug severity classification.
- **Hybrid Models:** A combination of different types of techniques of machine learning can potentially improve the performance. To get better trade-offs between precision and recall, it would be worth looking into a hybrid model that combines the interpretability of Logistic Regression with the power of deep learning models like DistilBERT.
- **Real-Time Prediction and Integration:** The development of real-time bug prediction systems that can dynamically prioritise bugs, as well as the addition of models to existing bug tracking systems like Bugzilla or JIRA to automate the processes of predicting bug severity and triaging bugs.
- **Model Explainability:** Although deep learning models such as DistilBERT provide a good level of performance, its black box nature makes it difficult to comprehend the model's predictions. Future research could potentially explore how these models can be made more explainable, in a manner that makes it more transparent and trusted by the developers.